\documentclass[10pt,aps,twocolumn,floats,floatfix,amssymb,prl,superscriptaddress]{revtex4-2}

\usepackage{graphicx}
\usepackage{amsmath,amssymb}
\usepackage{amsfonts}
\usepackage{xspace} 
\usepackage{xcolor}
\usepackage{dcolumn}
\usepackage{bm}
\usepackage{mathrsfs}
\definecolor{Burgundy}{RGB}{144,0,32}
\usepackage[colorlinks=true,linkcolor=Burgundy,citecolor=Burgundy, filecolor=Burgundy, urlcolor=Burgundy]{hyperref}
\usepackage[all]{hypcap} 
\usepackage[utf8]{inputenc} 
\usepackage{multirow}
\usepackage{etoolbox}
\usepackage{tikz}
\usepackage{fontawesome}

\newcommand{\dd}{\mathrm{d}}

\begin{document}

\title{
Leading effective field theory corrections to the Kerr metric at all spins
}

\author{Pedro G.~S.~Fernandes}
\email{fernandes@thphys.uni-heidelberg.de}
\affiliation{Institut f\"ur Theoretische Physik, Universit\"at Heidelberg, Philosophenweg 12, 69120 Heidelberg, Germany}

\begin{abstract}
The leading corrections to General Relativity can be parametrized by higher-derivative interactions in a low-energy effective field theory, in a way that is general and agnostic to the precise UV completion of gravity. Using numerical methods, we compute the leading-order corrections to the Kerr metric across the entire range of sub-extremal values of spin and analyse their impact on physical quantities. We find that rapidly rotating black holes are most affected by the higher-derivative corrections, making them especially sensitive probes of new physics. A dataset of solutions and the code used to produce them are publicly available.
\end{abstract}

\maketitle

\noindent \textbf{\textit{Introduction.}}
The existence of black holes (BHs) described by the Kerr metric~\cite{PhysRevLett.11.237} is a sharp prediction of General Relativity (GR), stemming from the uniqueness of the Kerr solution as the stationary, asymptotically flat vacuum BH solution of the theory~\cite{PhysRevLett.26.331,PhysRevLett.34.905}. With current and forthcoming gravitational-wave detections~\cite{LIGOScientific:2016aoc,LIGOScientific:2016lio,LIGOScientific:2019fpa,LIGOScientific:2020tif,LIGOScientific:2021sio,KAGRA:2021vkt,LIGOScientific:2025slb,2017arXiv170200786A,Barausse:2020rsu,TianQin:2015yph,10.1093/nsr/nwx116,Punturo:2010zz,LIGOScientific:2025obp} and very-long-baseline-interferometry observations~\cite{EventHorizonTelescope:2019dse,EventHorizonTelescope:2022wkp,EventHorizonTelescope:2022xqj,Ayzenberg:2023hfw,Johnson:2023ynn}, we are in a unique position to test this prediction with unprecedented precision.

Although GR has passed a wide range of experimental tests, it is best understood as the leading-order contribution in a low-energy effective field theory (EFT) that includes infinitely many higher-derivative operators~\cite{Donoghue:1995cz,Burgess:2003jk}.
A low-energy EFT encodes UV physics through a set of local interactions among the light degrees of freedom that respect the symmetries of the system, in a way that is general and agnostic to the precise UV completion.
In the simplest case, where the only light gravitational degrees of freedom are those of the metric and the fundamental symmetry is diffeomorphism invariance, the Einstein-Hilbert action receives corrections from higher-derivative operators built out of the Riemann tensor and its covariant contractions with the metric.
EFTs of gravity have been explored in many contexts~\cite{Cano:2019ore,deRham:2020ejn,Endlich:2017tqa,Melville:2024zjq,Lam:2025elw,Lam:2025fzi,Horowitz:2023xyl,Horowitz:2024dch,deRham:2021bll,deRham:2022gfe,Bern:2021ppb,Reall:2019sah,Ruhdorfer:2019qmk,Bernard:2025dyh,Cardoso:2018ptl,deRham:2019ctd,Allahyari:2025bvf,Figueras:2024bba,Davies:2023qaa,Xie:2021bur,Silva:2022srr,Li:2022pcy,Serra:2022pzl,Ma:2023qqj,Silva:2024ffz,Kehagias:2024yyp,Ma:2024ulp,Cao:2025qws,Cayuso:2023xbc,Corman:2024cdr,Figueras:2020ofh,Hollands:2022fkn,Miguel:2023rzp, Cano:2020cao,Cano:2021myl,Cano:2022wwo,Chung:2023zdq,Cano:2023tmv,Cano:2023jbk,Chung:2023wkd,Cano:2024wzo,Chung:2024ira,Chung:2024vaf,Cano:2024bhh,Cano:2024jkd,Cano:2024ezp,Maenaut:2024oci,Cano:2025zyk,Chung:2025gyg,Chung:2025wbg,Cano:2025mht,Cano:2025ejw,Franchini:2025csk,Blazquez-Salcedo:2024oek,DelPorro:2025fiu,Gies:2016con,Baldazzi:2023pep,Goon:2016mil}, including scenarios with extra light gravitational degrees of freedom which arise in certain UV completions of gravity, including string theory~\cite{Cano:2021rey,Sotiriou:2013qea,Sotiriou:2014pfa,Delgado:2020rev,Kleihaus:2011tg,Kleihaus:2015aje,Maselli:2015tta,Ayzenberg:2014aka,Yunes:2009hc,Yagi:2012ya,Lam:2025elw,Lam:2025fzi,Yunes:2009hc,Yagi:2012ya,Cardoso:2018ptl,Stein:2014xba,Konno:2014qua,Maselli:2015tta,Ayzenberg:2014aka,Chung:2024ira,Chung:2024vaf,Chung:2025gyg}.

When higher-derivative operators are considered, the Kerr metric acquires corrections.
Since the Kerr metric is Ricci-flat, most operators are redundant. At the four-derivative level, none of the operators contribute to leading-order corrections, as all non-trivial independent terms are at least quadratic in the Ricci tensor. At the six-derivative level, after accounting for field redefinitions~\footnote{Any operator proportional to the Ricci tensor can be eliminated by an appropriate field-redefinition of the metric.}, only two independent operators remain. These generate leading-order corrections to the Kerr metric. Thus, the most general EFT with up to six derivatives, that accounts for purely gravitational effects is~\cite{Cano:2019ore,deRham:2020ejn,Endlich:2017tqa,Melville:2024zjq}
\begin{equation}
    \begin{aligned}
        S = \frac{M_{\rm pl}^2}{2} \int \dd^4x \sqrt{-g} \bigg[ R &+ \frac{\lambda}{\Lambda^4} R_{\mu \nu}^{\phantom{\mu} \phantom{\nu} \rho \sigma} R_{\rho \sigma}^{\phantom{\rho} \phantom{\sigma} \delta \gamma} R_{\delta \gamma}^{\phantom{\delta} \phantom{\gamma} \mu \nu} \\ & + \frac{\tilde{\lambda}}{\Lambda^4} R_{\mu \nu}^{\phantom{\mu} \phantom{\nu} \rho \sigma} R_{\rho \sigma}^{\phantom{\rho} \phantom{\sigma} \delta \gamma} \tilde{R}_{\delta \gamma}^{\phantom{\delta} \phantom{\gamma} \mu \nu} \bigg],
        \label{eq:SEFT}
    \end{aligned}
\end{equation}
where $\tilde{R}_{\mu \nu \alpha \beta}$ is the dual-Riemann tensor, $\lambda$ and $\tilde{\lambda}$ are dimensionless Wilson coefficients, and $\Lambda$ denotes the energy scale that characterizes the regime of validity of the EFT which is expected to be directly related to the mass of the lightest field that was integrated out~\cite{Goon:2016mil}. The operator controlled by $\tilde{\lambda}$ is not parity-preserving, and therefore $\tilde{\lambda}$ would vanish in the case of a parity-preserving UV completion. While the sign of these Wilson coefficients is so far not constrained by positivity bounds~\cite{deRham:2022gfe,Bern:2021ppb}, asymptotically safe quantum gravity, for instance, predicts positive $\lambda$~\cite{DelPorro:2025fiu,Gies:2016con,Baldazzi:2023pep}.

Corrections to the Kerr metric arising from this EFT were computed in Ref.~\cite{Cano:2019ore}, perturbatively in a small-spin expansion. While the small-spin expansion yields accurate solutions for moderate spins, it breaks down for rapidly rotating BHs. Since observed populations contain BHs with large spin ~\cite{KAGRA:2021vkt,LIGOScientific:2025slb,Reynolds:2013rva}, there is a dire need to go beyond the small-spin expansion, and obtain accurate solutions at high spins.

Beyond the small-spin expansion, analytic solutions remain elusive due to the complex structure of the field equations. As a result, numerical methods become necessary. The construction of stationary, axisymmetric solutions through numerical techniques has seen significant progress in recent years~\cite{Fernandes:2022gde,Sullivan:2020zpf,Dias:2015nua}. Using these techniques, rapidly rotating BHs have been constructed in various modified theories of gravity with additional degrees of freedom, and in other settings~\cite{Delgado:2020rev,Kleihaus:2011tg,Kleihaus:2015aje,Delsate:2018ome,Herdeiro:2020wei,Berti:2020kgk,Eichhorn:2025aja,Cunha:2019dwb,Staykov:2025lfh,Fernandes:2024ztk,Collodel:2019kkx,Garcia:2023ntf,Guo:2023mda,Xiong:2023bpl,Liu:2025bkz,Cheng:2025hdw,Herdeiro:2025blx,Liu:2025eve,Lam:2025elw,Lam:2025fzi,Herdeiro:2014goa,Herdeiro:2016tmi,Burrage:2023zvk,Fernandes:2025osu,Destounis:2025tjn}.

In this Letter, we present the first accurate computation of the leading corrections to the Kerr metric from the EFT in Eq.~\eqref{eq:SEFT}, valid for all sub-extremal values of spin. We find that rapidly rotating BHs provide a highly sensitive window into new physics.
The solutions, together with the code used to generate them, are publicly available, allowing the community to readily use and build upon them.
We work in units $c=G=1$.

\noindent \textbf{\textit{Corrections to the Kerr metric.}}
The Kerr metric in Boyer-Lindquist coordinates is given by
\begin{equation}
    \begin{aligned}
        \dd s^2_{(0)} =& -\frac{\Delta}{\Sigma} \left( \dd t - a \sin^2 \theta \dd \varphi\right)^2 + \Sigma \left( \frac{\dd r^2}{\Delta} + \dd \theta^2 \right)\\ & + \frac{\sin^2 \theta}{\Sigma} \left( a \dd t - (r^2+a^2) \dd \varphi \right)^2,
    \end{aligned}
\end{equation}
where $M$ and $a$ are, respectively, the Arnowitt-Deser-Misner mass and angular momentum per unit mass of the BH, $\Sigma = r^2+a^2\cos^2\theta$, and $\Delta = r^2 - 2M r+a^2$. This background metric is Ricci-flat, $R_{\mu \nu}^{(0)} = 0$.

We consider perturbative corrections to the Kerr metric in the EFT framework. These are obtained by solving the Einstein field equations to leading-order in the expansion parameter $\varepsilon \equiv 1/(M \Lambda)^4 \ll 1$~\footnote{Reinstating $G$, the expansion parameter becomes $\varepsilon\equiv 1/(GM \Lambda)^4 \sim M_{\rm pl}^8/(M\Lambda)^4$.}. Expressing the corrections as
\begin{equation}
    g_{\mu \nu} = g_{\mu \nu}^{(0)} + \varepsilon g_{\mu \nu}^{(1)},
\end{equation}
the Einstein tensor becomes $G_{\mu \nu} = \varepsilon G_{\mu \nu}^{(1)}$ to leading order, where
\begin{equation}
    G_{\mu \nu}^{(1)} = -\frac{1}{2} \nabla^2 \hat{g}_{\mu \nu}^{(1)} - \frac{1}{2} g_{\mu \nu}^{(0)} \nabla^\alpha \nabla^\beta \hat{g}_{\alpha \beta}^{(1)} + \nabla^\alpha \nabla_{(\mu} \hat{g}_{\nu) \alpha}^{(1)},
\end{equation}
and where we defined the trace-reversed metric correction
\begin{equation}
    \hat{g}_{\mu \nu}^{(1)} = g_{\mu \nu}^{(1)} - \frac{1}{2} g_{\mu \nu}^{(0)} g_{\alpha \beta}^{(1)} g^{(0)\alpha \beta}.
\end{equation}
The Einstein field equations determining leading-order corrections to the Kerr metric become
\begin{equation}
    G_{\mu \nu}^{(1)} = M^4 \left( \lambda T_{\mu \nu}^{(\mathrm{ev})} + \tilde{\lambda} T_{\mu \nu}^{(\mathrm{odd})} \right)\bigg|_{g=g^{(0)}},
    \label{eq:field_eqs_pert}
\end{equation}
where the stress-energy tensors associated with the parity-even and parity-odd operators, whose expressions are given in Ref.~\cite{Cano:2019ore} and in the Supplemental Material, are evaluated using the Kerr metric. They are the source for the metric corrections.
Since the field equations are linear in the corrections $g_{\mu \nu}^{(1)}$, the system can be solved independently for the parity-even and parity-odd sectors by decomposing the metric
\begin{equation}
    g_{\mu \nu}^{(1)} = \lambda g_{\mu \nu}^{(1,\mathrm{ev})} + \tilde{\lambda} g_{\mu \nu}^{(1,\mathrm{odd})}.
    \label{eq:even_odd_split}
\end{equation}

Following Ref.~\cite{Cano:2019ore}, we parametrize corrections to the Kerr metric in terms of four functions, $H_i (r,\theta)$ where $i \in \{1,2,3,4\}$, as
\begin{equation}
    \begin{aligned}
        \dd s^2_{(1)} =& H_1 \dd t^2 - H_2\frac{4M a r \sin^2\theta}{\Sigma}\dd t \dd \varphi + H_3 \Sigma \left( \frac{\dd r^2}{\Delta} + \dd \theta^2 \right)\\& 
        + H_4 \left( r^2+a^2 + \frac{2M a^2 r \sin^2\theta}{\Sigma} \right)\sin^2\theta \dd \varphi^2.
    \end{aligned}
\end{equation}
This is the most general parametrization (up to diffeomorphisms and redefinitions), since our framework falls under the assumptions of the theorem presented in Ref.~\cite{Xie:2021bur}.
These functions are split into parity-even and parity-odd sectors as in Eq. \eqref{eq:even_odd_split}, $H_i = \lambda H_i^{\rm (ev)} + \tilde{\lambda} H_i^{\rm (odd)}$. This parametrization has the advantage that the coordinate location of the event horizon is always the same as for the Kerr metric, $r_h = M+\sqrt{M^2-a^2}$.

To ensure the metric is asymptotically flat, and $M$ and $a$ retain their physical meaning, the metric corrections obey the following boundary conditions~\cite{Cano:2019ore}
\begin{equation}
    \begin{aligned}
        &\lim_{r\to \infty} H_1 = 0, \quad \lim_{r\to \infty} H_2 = -\lim_{r\to \infty} H_3/2,\\
        & \lim_{r\to \infty} H_3 = \lim_{r\to \infty} H_4 = \lim_{r\to \infty} r^2 \partial_r H_3 / M.
    \end{aligned}
    \label{eq:bcs}
\end{equation}

The field equations \eqref{eq:field_eqs_pert} constitute a system of four independent second-order \emph{linear} partial differential equations in the radial and angular coordinates $r$ and $\theta$, for the functions $H_i$.
This linear structure allows the system to be reformulated as a linear algebra problem, greatly simplifying the solution-finding procedure. A similar situation was encountered in Refs.~\cite{Lam:2025elw,Lam:2025fzi} by treating scalar-Gauss-Bonnet~\cite{Sotiriou:2013qea,Sotiriou:2014pfa,Delgado:2020rev,Kleihaus:2011tg,Kleihaus:2015aje,Maselli:2015tta,Ayzenberg:2014aka} and dynamical Chern-Simons~\cite{Yunes:2009hc,Yagi:2012ya} theories as EFTs.

To solve the field equations, we use a pseudospectral collocation method~\cite{Fernandes:2022gde}, where each of the functions $H_i$ is expanded in a spectral series in the coordinates $x=1-2r_h/r$, and $y=\cos\theta$, with resolutions $N_x$ and $N_y$, respectively. The reader is referred to the Supplemental Material for a detailed description of the numerical method.

\noindent \textbf{\textit{Results.}}
A main result of this Letter is the successful solution of the field equations within this EFT framework, yielding a comprehensive dataset of corrections to the Kerr metric, for both parity sectors, for all spins from $a/M = 0$ to $a/M = 0.99$ in increments of $0.01$, as well as near-extremal solutions at $a/M = 0.999$. This dataset, along with the code used to generate the solutions, is publicly available.

\begin{figure}[]
	\centering
	\includegraphics[width=\linewidth]{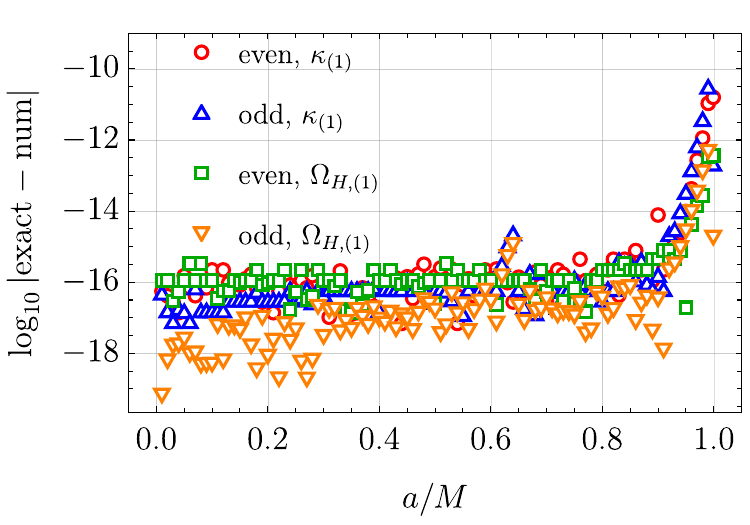}\hfill
    \caption{Absolute difference between the numerical and exact values of $\kappa_{(1)}$ and $\Omega_{H,(1)}$ for both parity sectors. The numerical values were obtained using resolutions $(N_x, N_y) = (50,16)$ for $a/M < 0.65$, $(N_x, N_y) = (60,24)$ for $0.65 \leq a/M < 0.999$, and $(N_x, N_y) = (64,28)$ for $a/M = 0.999$. In the parity-odd case, the exact values of these quantities are identically zero.}
	\label{fig:diffs}
\end{figure}

\begin{figure*}[]
	\centering
	\includegraphics[width=0.5\linewidth]{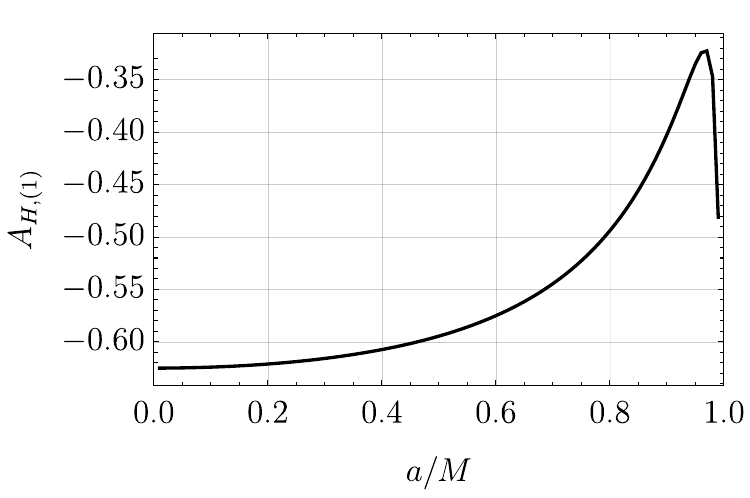}\hfill
	\includegraphics[width=0.5\linewidth]{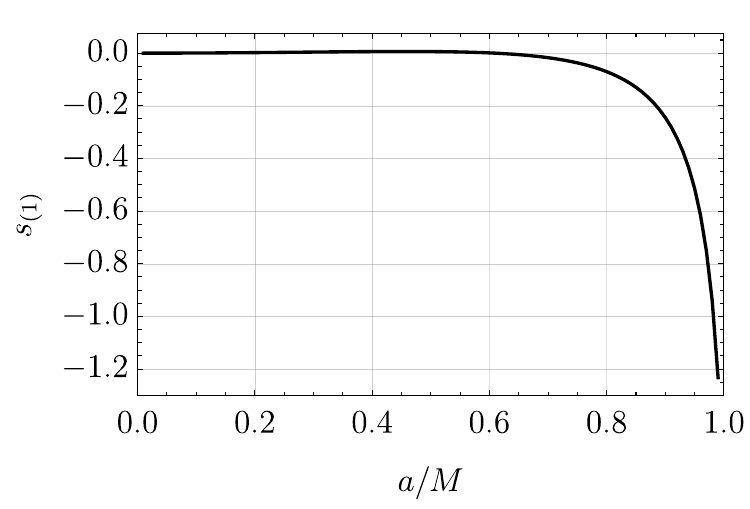}\vfill
	\includegraphics[width=0.5\linewidth]{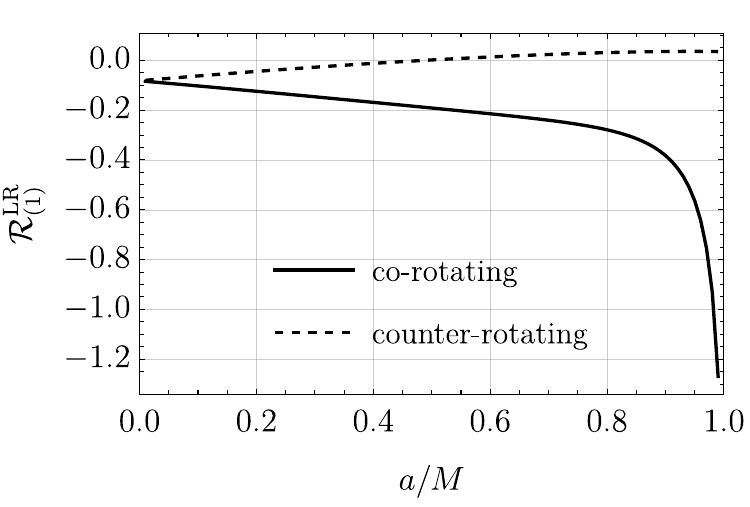}\hfill
	\includegraphics[width=0.5\linewidth]{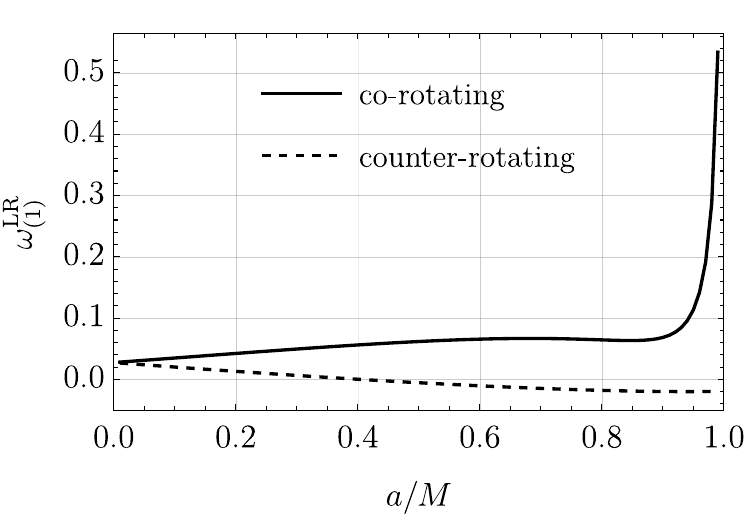}\vfill
    \caption{Corrections to the area of the horizon (top left), sphericity (top right), perimetral location of the light ring (bottom left) and orbital frequency at the light ring (bottom right) for spin values up to $a/M=0.99$.}
	\label{fig:quantities}
\end{figure*}

\begin{figure}[]
	\centering
	\includegraphics[width=\linewidth]{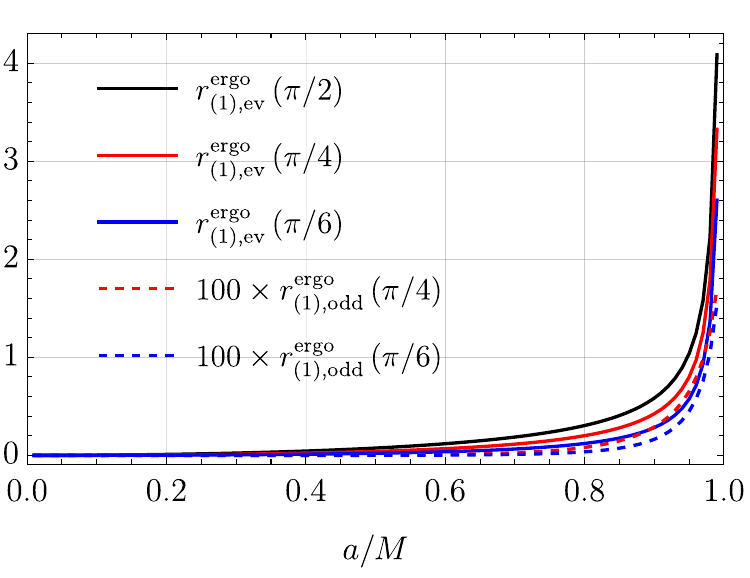}\hfill
    \caption{Corrections to the location of the ergosphere for several values of $\theta$, shown for both parity-even and parity-odd corrections.}
	\label{fig:ergo}
\end{figure}

To evaluate the accuracy of our solutions, we have performed several convergence tests. We observed exponential convergence with increasing resolutions, and excellent agreement with the small-spin expansion~\cite{Cano:2019ore} for small values of spin, which we discuss in detail below. Since analytic expressions for the EFT corrections to certain BH thermodynamic quantities were derived in Ref.~\cite{Reall:2019sah} (without requiring explicit computation of the metric corrections), we compare our numerically obtained corrections to the surface gravity $\kappa$ and the horizon angular velocity $\Omega_H$ with the corresponding analytic results from Ref.~\cite{Reall:2019sah}. Throughout this work, corrections to physical quantities are written as
\begin{equation}
    X = X_{(0)} \left( 1 + \varepsilon \lambda\, X_{\mathrm{ev},(1)}+ \varepsilon \tilde{\lambda}\, X_{\mathrm{odd},(1)} \right),
\end{equation}
where $X$ denotes an arbitrary physical quantity. This parametrization allows us to characterize the corrections to physical quantities in a way that is independent of the specific values of $\varepsilon$, $\lambda$, and $\tilde{\lambda}$.
The formulas used to compute the corrections to the physical quantities considered in this work, in terms of the metric functions $H_i$, are provided in the Supplemental Material. In Fig.~\ref{fig:diffs}, we show the absolute difference between the numerically computed values of $\kappa_{(1)}$ and $\Omega_{H,(1)}$ and their exact values, plotted as a function of the dimensionless spin $a/M$. We find that for spins $a/M \lesssim 0.9$, the errors remain below machine precision, although they increase as the spin approaches extremality. Even for near-extremal BHs, however, the errors stay below $\mathcal{O}\left(10^{-10}\right)$ at the resolutions considered, which is sufficient for most practical applications. If higher accuracy is needed, it can be achieved by increasing the numerical resolution.

As an example of how these numerical solutions make it possible to compute physically meaningful quantities that would otherwise be inaccessible, we have computed the impact of the EFT corrections on the horizon area $A_H$, the BH sphericity $s$, the location and geodesic frequency of the light rings, and the location of the ergosphere. The sphericity of the BH is defined as the ratio between the circumference of the horizon measured along the equator and that measured along the poles. If $s>1$, as in the case of a rotating BH in GR, the horizon has the shape of an oblate spheroid.
Light rings are circular orbits of massless particles confined to the equatorial plane. They are directly related to the BH shadow~\cite{Cunha:2018acu}, making them particularly relevant observationally, since very-long-baseline interferometry can, in principle, probe EFT corrections to quantities associated with the light ring.
As discussed in Ref.~\cite{Cano:2019ore}, these orbits cease to exist when parity-violating terms are included; therefore, we set $\tilde{\lambda}=0$, throughout the analysis of the light rings. Since the coordinate radius of the light ring is not a physically meaningful quantity, we present our results in terms of the perimetral radius, defined as $\mathcal{R} = \sqrt{g_{\varphi \varphi}} \big|_{\theta=\pi/2}$.

The results are presented in Fig.~\ref{fig:quantities} and Fig.~\ref{fig:ergo}. In agreement with Ref.~\cite{Cano:2019ore}, we find that parity-odd corrections do not affect quantities evaluated at the horizon, such as the horizon area $A_H$ and the sphericity $s$, and are therefore not shown in Fig.~\ref{fig:quantities}. They do, however, modify quantities such as the location of the ergosphere, cf. Fig.~\ref{fig:ergo}, although their magnitude shows that these parity-odd corrections are subleading.

The corrections to the horizon area are not monotonic with increasing spin, contrary to the prediction of the small-spin expansion~\cite{Cano:2019ore}. Instead, we find that for very large spins, $A_{H,(1)}$ becomes increasingly negative. The values of $s_{(1)}$ grow rapidly at large spin, driving the horizon geometry towards prolateness for positive $\lambda$ and towards oblateness for negative $\lambda$. Positive (negative) $\lambda$ shifts the ergosphere to a larger (smaller) radius.
We do not display deviations for spins larger than $a/M=0.99$, as their magnitude becomes extremely large, following the trend observed in Fig.~\ref{fig:quantities}.

Concerning the light ring, we find that the corrections are larger for co-rotating orbits than for counter-rotating ones. This is attributed to the fact that co-rotating orbits lie closer to the horizon and thus probe deeper into the strong-field region, where the EFT corrections are larger. Generally, higher-curvature corrections make the photon sphere more (less) compact for positive (negative) $\lambda$.

Since the magnitude of the corrections exhibits rapid growth as the extremal limit is approached, rapidly rotating BHs are found to be powerful amplifiers of new physics, in agreement with the results of Ref.~\cite{Horowitz:2023xyl}.

\begin{figure}[h!]
	\centering
	\includegraphics[width=\linewidth]{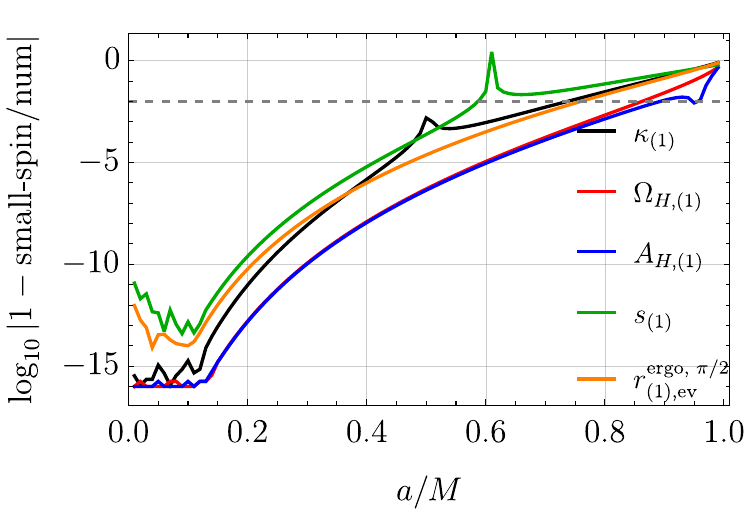}\hfill
    \caption{Relative differences between the values computed using the small-spin expansion up to $\mathcal{O}\left((a/M)^{15}\right)$ and the numerical results for various physical quantities.}
	\label{fig:accuracy_small_spin}
\end{figure}

\noindent \textbf{\textit{Breakdown of the small-spin expansion.}}
To evaluate the accuracy of the small-spin expansion derived in Ref.~\cite{Cano:2019ore}, we compare the numerically computed EFT corrections with the corresponding results from the small-spin expansion truncated at $\mathcal{O}\left((a/M)^{15}\right)$. The relative differences are displayed in Fig.~\ref{fig:accuracy_small_spin}.

We find that the accuracy of the small-spin expansion depends somewhat on the specific physical quantity under consideration, although it consistently breaks down at sufficiently large spin. In all quantities, the relative error exceeds the percent level for spins $a/M \gtrsim 0.85$, while for some quantities, such as the sphericity, this threshold is reached already at $a/M \gtrsim 0.6$.

These results underscore the importance of the numerical solutions: for physical predictions at large spin, they are essential.
Our results therefore open the door to exploring the large-spin regime.

\noindent \textbf{\textit{Discussion.}}
In this Letter, we have employed pseudospectral methods to solve the system of partial differential equations governing the leading-order corrections to the Kerr metric within the EFT framework of Eq.~\eqref{eq:SEFT}, and to compute the corresponding leading-order corrections to physical quantities of interest.
Both the code developed to solve the field equations and the accompanying dataset of solutions are publicly available.

We find that rapidly rotating BHs strongly amplify new physics, as all computed EFT corrections to physical quantities grow rapidly as extremality is approached. Moreover, as expected, the small-spin expansion breaks down at large spins, highlighting the importance of the numerical solutions presented here for accurately probing the large-spin regime of EFT corrections to the Kerr metric.

The methods employed in this Letter can be generalized to compute corrections involving eight or more derivatives, which provide the leading-order corrections to the Kerr metric in certain cases, such as when the UV completion is supersymmetric~\cite{Camanho:2014apa}, or when the EFT is isospectral~\cite{Cano:2024wzo}. These are, however, extremely special cases, and the leading corrections to the Kerr metric are expected to be those studied in this Letter.

Looking ahead, these numerical solutions open the door to computing the quasi-normal modes of EFT-corrected, rapidly rotating BHs, extending the analyses of Refs.~\cite{Cano:2020cao,Cano:2021myl,Cano:2023jbk,Cano:2024ezp,Maenaut:2024oci}. This direction is crucial for testing GR with gravitational-wave observations, since such observations may in principle be sensitive to EFT corrections to the quasi-normal mode spectrum, and, as our results indicate, rapidly rotating BHs are especially powerful probes of new physics. Work along these lines is in progress, and the results will be reported in a forthcoming publication.

\noindent \textbf{\textit{Acknowledgments.}}
P.F. thanks Kelvin Lam for useful correspondence, and Bruno Valeixo Bento, Astrid Eichhorn, Benjamin Knorr and Simon Maenaut for valuable comments on a first draft.
This work is funded by the Deutsche Forschungsgemeinschaft (DFG, German Research Foundation) under Germany’s Excellence Strategy EXC 2181/1 - 390900948 (the Heidelberg STRUCTURES Excellence Cluster).

\noindent \textbf{\textit{Data availability.}} The code used to solve the field equations, along with a dataset of solutions, is available at \href{https://github.com/pgsfernandes/EFT-Corrections-Kerr-Metric}{this repository \faGithub}.

\bibliographystyle{apsrev4-1}
\bibliography{References}

\onecolumngrid

\renewcommand{\theequation}{SM.\arabic{equation}}
\setcounter{equation}{0}
\section*{Supplemental Material}
\subsection*{Stress-energy tensors associated with the higher-derivative corrections}
The stress-energy tensors associated with the higher-derivative corrections, that appear as the source term for the metric corrections in Eq.~\eqref{eq:field_eqs_pert}, are given by~\cite{Cano:2019ore}
\begin{equation}
    T_{\mu \nu}^{(\mathrm{ev})} = 3 R_{\mu}^{\phantom{\mu}\sigma \alpha \beta} R_{\alpha \beta}^{\phantom{\alpha \beta}\rho \lambda} R_{\rho \lambda \sigma \nu} + \frac{1}{2} g_{\mu \nu} R_{\alpha \beta}^{\phantom{\alpha \beta}\rho \sigma} R_{\rho \sigma}^{\phantom{\alpha \beta}\delta \lambda} R_{\delta \lambda}^{\phantom{\delta \lambda}\alpha \beta} - 6 \nabla^\alpha \nabla^\beta \left( R_{\mu \alpha \rho \lambda} R_{\nu \beta}^{\phantom{\nu \beta}\rho \lambda} \right),
\end{equation}
\begin{equation}
    \begin{aligned}
        T_{\mu \nu}^{(\mathrm{odd})} =& -\frac{3}{2} R_{\mu}^{\phantom{\mu}\rho \alpha \beta} R_{\alpha \beta \sigma \lambda} \tilde{R}_{\nu \rho}^{\phantom{\nu \rho}\sigma \lambda} - \frac{3}{2} R_{\mu}^{\phantom{\mu}\rho \alpha \beta} R_{\nu \rho \sigma \lambda} \tilde{R}_{\alpha \beta}^{\phantom{\alpha \beta} \sigma \lambda} + \frac{1}{2} g_{\mu \nu} R_{\alpha \beta}^{\phantom{\alpha \beta}\rho \sigma} R_{\rho \sigma}^{\phantom{\alpha \beta}\delta \lambda} \tilde{R}_{\delta \lambda}^{\phantom{\delta \lambda}\alpha \beta} \\& + 3 \nabla^\alpha \nabla^\beta \left( R_{\mu \alpha \sigma \lambda} \tilde{R}_{\nu \beta}^{\phantom{\nu \beta}\sigma \lambda} + R_{\nu \beta \sigma \lambda} \tilde{R}_{\mu \alpha}^{\phantom{\mu \alpha}\sigma \lambda} \right).
    \end{aligned}
\end{equation}

\subsection*{Solving the field equations: the numerical method}

To solve the field equations, we use a pseudospectral collocation method that transforms the system into a linear algebra problem. In this approach, each of the metric corrections is expanded in a spectral series
\begin{equation}
    \begin{aligned}
        &H_i^{\rm (ev)}(x,y) = \sum_{n=0}^{N_x-1} \sum_{m=0}^{N_y-1} c_{i,n m} T_n(x) T_{2m}(y),\\&
        H_i^{\rm (odd)}(x,y) = \sum_{n=0}^{N_x-1} \sum_{m=0}^{N_y-1} \tilde{c}_{i,n m} T_n(x) T_{2m+1}(y),
    \end{aligned}
    \label{eq:spectral_series}
\end{equation}
where $T_k$ denotes the $k^{\rm th}$ order Chebyshev polynomial, the coefficients of the expansion are known as \emph{spectral coefficients}, $N_x$ and $N_y$ are the spectral resolutions in $x$ and $y$, respectively, and we have defined the coordinates
\begin{equation}
    x=1-\frac{2r_h}{r},\qquad y=\cos \theta,
\end{equation}
which are adapted to the numerical method. Their ranges are $x,y\in [-1,1]$. With these spectral series, the functions have definite parity, and allow us to consider only the range $y\in [0,1]$. In the following, we outline the procedure for one of the parity sectors, noting that the treatment of the other sector is analogous.

The left-hand-side of the perturbed Einstein equations \eqref{eq:field_eqs_pert} has four independent components, which we take to be the $(tt)$, $(rr)$, $(\theta \theta)$ and $(t\varphi)$ components. After employing the spectral series, these field equations depend on $x$ and $y$, and have $4 \times N_x \times N_y$ unknowns corresponding to the spectral coefficients. The right-hand-side consists of a source term that depends solely on $x$ and $y$. To close the system, the number of equations must match the number of unknown spectral coefficients. This is achieved by evaluating the four independent field equations at all combinations of collocation points given by
\begin{equation}
    \begin{aligned}
        x_n = \cos \left( \frac{n}{N_x}\pi \right), \quad &n=0,\ldots,N_x-1,\\
        y_n = \cos \left( \frac{2n+1}{4N_y}\pi \right), \quad &n=0,\ldots,N_y-1,
    \end{aligned}
\end{equation}
forming a two-dimensional grid with $N_x\times N_y$ points. The grid in $y$ does not include the boundary points $y=0$ and $y=1$, as no explicit boundary conditions are required there. The spectral series automatically captures the correct behaviour of the metric functions at these boundaries. In contrast, the grid in $x$ includes one boundary point, $x=1$, corresponding to spatial infinity, where the boundary conditions given in Eq.~\eqref{eq:bcs} are imposed (instead of evaluating the differential equations at that point).
Although the field equations are second-order, no boundary conditions need to be applied at $x=-1$. This can be understood as follows: the general solution to the field equations can be decomposed into a homogeneous and a particular part, with the boundary conditions fixing the homogeneous component, which depends on two integration constants. The conditions in Eq.~\eqref{eq:bcs} select the homogeneous solution that ensures asymptotic flatness and preserves the physical interpretation of $M$ and $a$. The remaining terms in the homogeneous solution are non-regular at the horizon, exhibiting, for instance, divergent derivatives there~\cite{Cano:2019ore,Lam:2025fzi}. Since the spectral series in Eq.~\eqref{eq:spectral_series} are, by construction, regular at the horizon, they inherently exclude these non-regular contributions.

Following this procedure, the system of differential equations is transformed into a matrix equation of the form $A \mathbf{v} = \mathbf{b}$, where $\mathbf{v}$ is a column vector containing all the spectral coefficients. The matrix $A$ represents the Jacobian of the system, with dimensions $(4 \times N_x \times N_y) \times (4 \times N_x \times N_y)$, obtained by differentiating the field equations on the grid with respect to all spectral coefficients. The vector $\mathbf{b}$ has $4 \times N_x \times N_y$ components and corresponds to the source term evaluated at each point on the grid. This system can be solved using any linear-algebra library, for each value of the spin parameter $a/M \in [0,1)$. We have used the \texttt{LinearSolve.jl} and \texttt{DoubleFloats.jl} libraries of \texttt{Julia} to solve the system with high-precision. In the numerical method we fix $M=1$, without loss of generality.

\subsection*{Formulas to Compute EFT Corrections to Physical quantities}
To compute the EFT corrections to the surface gravity $\kappa$, horizon angular velocity $\Omega_H$, and area of the Horizon $A_H$, we use the expressions obtained in Ref.~\cite{Cano:2019ore}
\begin{equation}
    \kappa_{(1)} = \left[ H_2 - \frac{H_3+H_4}{2} + M^2 r_h^2 \frac{\partial_r \left(- H_1 \Sigma + a^2 (1-y^2) (2H_2-H_4) \right) + 2(r_h-M)(H_4-2H_2)}{(r_h-M)(r_h^2 + a^2 y^2)^2} \right]_{r=r_h},
\end{equation}

\begin{equation}
    \Omega_{H,(1)} = \left[ H_2 - H_4 \right]_{r=r_h},
\end{equation}

\begin{equation}
    A_{H,(1)} = \frac{1}{4} \int_{-1}^{1} \left[ H_3 + H_4 \right]_{r=r_h} \dd y.
\end{equation}

The sphericity $s$ is defined as the ratio of the circumference of the horizon measured along the equator
\begin{equation}
    L_e = 2\pi \sqrt{g_{\varphi \varphi}}|_{r=r_h},
\end{equation}
and that measured along the poles
\begin{equation}
    L_p = 2 \int_{0}^{\pi} \sqrt{g_{\theta \theta}}|_{r=r_h} \dd \theta.
\end{equation}
Using these expressions, we find
\begin{equation}
    \begin{aligned}
        & L_{e,(1)} = \frac{1}{2} H_4|_{r=r_h}\\&
        L_{p,(1)} = \frac{1}{L_{p,(0)}} \int_{-1}^{1} \sqrt{\frac{r_h^2 + a^2 y^2}{1-y^2}} H_3 \dd y \rvert_{r=r_h}\\&
        s_{(1)} = L_{e,(1)}-L_{p,(1)},
    \end{aligned}
\end{equation}
where
\begin{equation}
    L_{p,(0)} = 4\sqrt{r_h^2 + a^2} E\left( \frac{a^2}{r_h^2+a^2} \right),
\end{equation}
and where $E(k)$ is the complete elliptic integral of second kind: $E(k) = \int_{0}^{\pi/2} \sqrt{1-k \sin^2 \theta} \dd \theta$.

The ergosphere is located where $g_{tt}=0$. Solving this condition perturbatively, we find
\begin{equation}
    r^{\rm ergo}_{(1)}(y) = \frac{1}{\sqrt{1-(a/M)^2 y^2}} H_1 \bigg|_{r=r^{\rm ergo}_{(0)}}.
\end{equation}

Following Ref.~\cite{Fernandes:2022gde}, the equation of motion for the radial coordinate of an equatorial geodesic is given by $\dot{r}^2 = V_{\rm eff}$, where
\begin{equation}
    V_{\rm eff} = \frac{E^2g_{\varphi \varphi} + 2 E L g_{t\varphi} + L^2 g_{tt}}{g_{rr} \left( g_{t\varphi}^2-g_{tt}g_{\varphi \varphi} \right)},
    \label{eq:Veff}
\end{equation}
is an effective potential, and where $E = -g_{tt} \dot{t} - g_{t\varphi} \dot{\varphi}$, and $L = g_{t\varphi} \dot{t} + g_{\varphi \varphi} \dot{\varphi}$, denote the specific energy and angular momentum of a massless test particle, respectively. Circular orbits are determined by the conditions $\dot{r}=0$ and $\ddot{r}=0$, which can be expressed in terms of the effective potential and its radial derivative. For non-zero BH spin, two distinct light rings exist: one co-rotating and one counter-rotating relative to the BH. The corresponding orbital frequencies are obtained from $\omega = \dot{\varphi}/\dot{t}$ evaluated at the light ring locations.

We note that if matter is minimally coupled in the frame where all higher-curvature operators are present, prior to their removal via field redefinitions, then these operators can influence physical observables that are not purely gravitational, such as quantities associated with the light ring and ergosphere. At the six-derivative parity-conserving level, there is a single additional operator that corrects the Kerr metric, and influences observables that depend on matter~\cite{Cano:2019ore,Allahyari:2025bvf}. It takes the form takes the form $\frac{c}{\Lambda^4} R_{\mu\nu\alpha\beta} R^{\mu\nu\alpha\beta} R$, and can be removed from the action by a field redefinition $g_{\mu\nu} \rightarrow \left(1 - \frac{c}{\Lambda^4} R_{\kappa\gamma\alpha\beta} R^{\kappa\gamma\alpha\beta}\right) g_{\mu\nu}$, thereby recovering the EFT of Eq. \eqref{eq:SEFT} at the expense of introducing a conformal coupling for test particles, where the conformal factor depends on the Kretschmann scalar. The contributions of order $\varepsilon$ from this operator to light-ring observables can be computed solely from the background Kerr geometry and its Kretschmann scalar, without requiring explicit calculation of the corresponding metric corrections, as was necessary for the operators in Eq. \eqref{eq:SEFT}. The coordinate location and orbital frequency of the light ring remain invariant under this field redefinition of the metric, whereas its perimetral location does not and must be computed explicitly.

\end{document}